\documentclass[9pt,twocolumn,twoside]{opticajnl}
\journal{opticajournal} 

\setboolean{shortarticle}{false}

\usepackage{listings}
\usepackage{xcolor}
\usepackage[T1]{fontenc}
\usepackage[utf8]{inputenc}      
\usepackage{listings}
\usepackage{listingsutf8}        

\usepackage{subcaption} 
\usepackage{lineno}
\usepackage{mathtools} 
\title{Diffractive Retroreflector for Distributed Sensing}

\author[1,*]{Anne R. Kroo}
\author[1]{Olav Solgaard}

\affil[1]{Department of Electrical Engineering, Stanford University, 348 Via Pueblo, Stanford, CA, 94305}

\affil[*]{akroo@stanford.edu}

\begin{abstract}
We introduce a modified corner cube reflector that encodes information from passive optical sensors in its retroreflected diffraction pattern, enabling remote sensor-state measurement over a single-ended optical link. The design interferes a reference path and a sensor-modulated path within the retroreflected beam to produce an interferometric signal suitable for reading out phase and amplitude variations with a square-law camera. This enables sensor-state determination in arbitrarily oriented passive nodes, extending coherent interferometric read out of chemical, biological, and physical sensors to scalable, robust, and inert field deployments.
\end{abstract}

\setboolean{displaycopyright}{false} 

\begin{document}

\maketitle

\section{Introduction}

Corner cubes, and other retroreflectors, reflect light back to distant light sources regardless of the incident angle, which allows them to enhance visibility, improve alignment, and increase measurement accuracy in a number of applications, spanning traffic safety, surveying, laser ranging distance measurements, barcode scanning, and remote identification. Corner cube retroreflectors (CCRs) have also been used for sensing and communication \cite{pister_micro_ccr_coms, olav_modulator_ccr} by modulating or angularly displacing one facet of the corner cube. Here we introduce a CCR with an integrated sensor, whose state strongly modifies the diffraction pattern from the CCR, allowing the sensor state to be remotely interrogated.

\begin{figure}[H]
\centering\includegraphics[width=\linewidth]{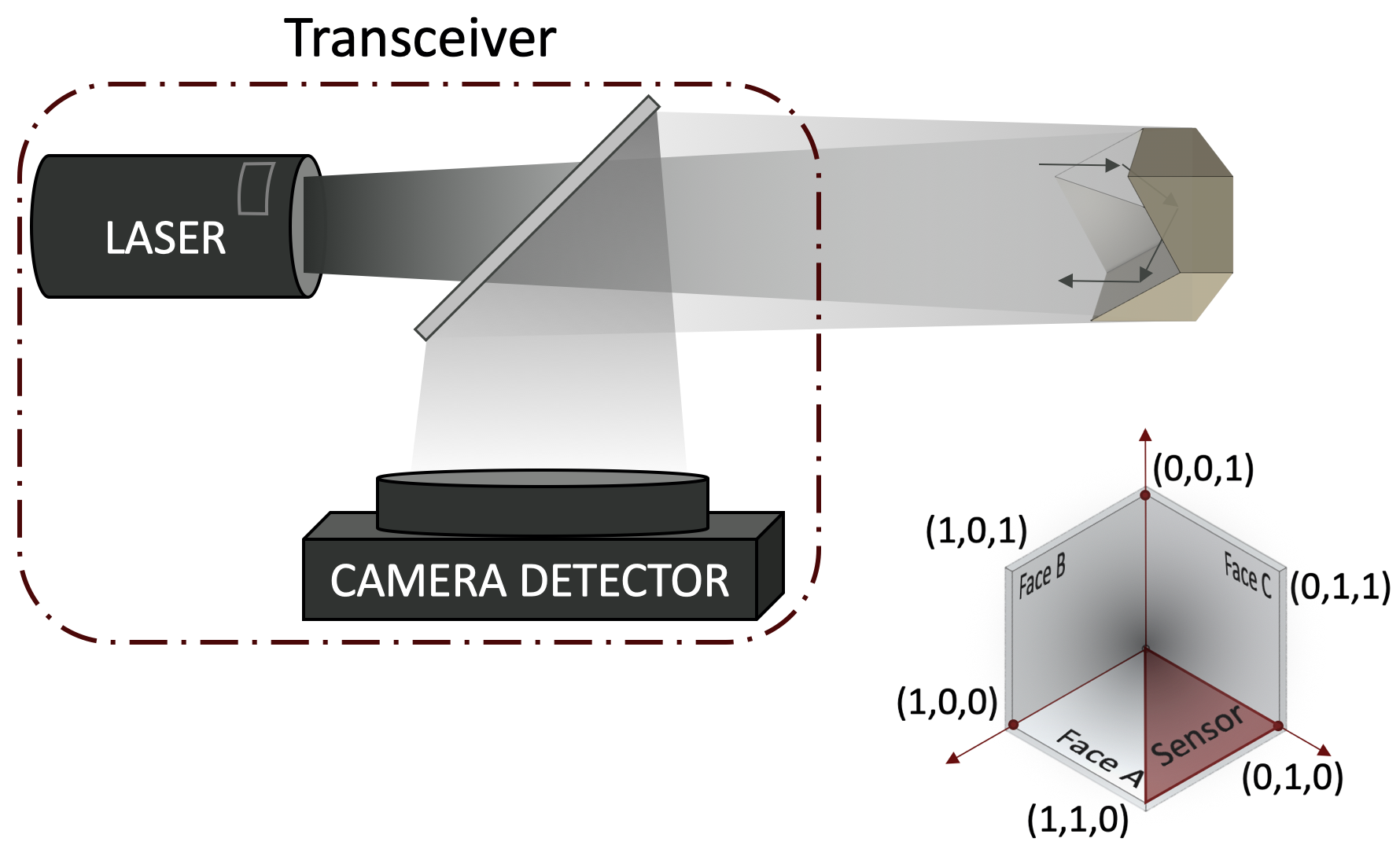}
\caption{System design of distributed sensor system with transceiver architecture and remote corner cube sensor and close up of corner cube with embedded sensor with corresponding coordinate system.}
\label{fig:system_design}
\end{figure}
Our corner cube consists of three mutually orthogonal square surfaces with an integrated phase or amplitude sensor. As shown in Fig. \ref{fig:system_design}, the sensor occupies one half of one of the three facets of the CCR. This placement leads to strong changes in the diffraction pattern as a function of the state of the sensor, as we show below. The sensor can be any material or patterning that changes its phase and/or reflectivity as a function of a measurand. To read the sensor information, we use a transceiver consisting of a laser and a camera. The laser illuminates the aperture of the corner cube, and light is retroreflected back to the transceiver. This returning diffraction pattern is captured by the camera. As the phase and/or reflectivity of the sensor surface varies relative to the reference surfaces, the diffraction pattern undergoes corresponding changes under coherent illumination. By leveraging robustness of interferometric readout, our approach offers a scalable, low-cost, self-referencing solution for distributed sensing. This paper outlines the design and simulations of this sensing paradigm, demonstrating its potential for large-scale deployment.
\section{Methods}
\subsection{Methods Overview}
For our applications, the dimensions of the system of Fig. \ref{fig:system_design} range from the millimeter scale for the CCR to tens or even hundreds of meters for the distance from the transceiver to the CCR. These facts allow us to make the following simplifying assumptions: 1) There is negligible diffraction within the CCR, so we model it with ray tracing. 2) the interrogating beam is flat across the aperture. 3) the interrogator is in the far field such that Fraunhofer Diffraction applies, and 4) the reflectors in the CCR do not impart polarization rotation, a good assumption for most mirrors except those based on total internal reflection. A standard hollow corner cube has a flat reflected near field \cite{scholl_ray_1995, eckhardt_simple_1971}, but this is not the case when we modulate part of the reflections as shown in Fig. \ref{fig:system_design}, because the near field at the output of the CCR will have some parts that have interacted with the modulator and some that have not.
We use ray tracing to identify how different portions of the retroreflected light have interacted with the embedded sensor and represent the decreasing effective aperture with angular rotation, yielding a complex two-dimensional near field. By applying the Fraunhofer Diffraction integral separately to the sensor or transducer influenced region (\textbf{\textit{T}}) and the non-sensor region (\textbf{\textit{N}}), we find the total diffraction pattern by summing the two contributions:
\begin{equation}D(X,Y,r,\theta) = r e^{i \theta}\oint_T e^{\frac{i k (X x + Y y)}{R}}dx dy+\oint_N e^{\frac{i k (X x + Y y)}{R}}dx dy
\label{eq:diff_integral}
\end{equation}

where $k = \frac{2\pi}{\lambda}$, $R$ is the propagation distance, $r$ is the reflectivity value of the sensor, and $\theta$ is the phase shift of the sensor.
Equation 1 has only two terms, because even though the amplitude and/or phase modulation that takes place in the sensor is angle sensitive, the incident angle onto the sensor is the same whether the light rays hit it on the first, second, or third bounce within the CCR. Intuitively this can be understood by considering the fact that at the three surfaces are orthogonal so each ray must get the same angular rotation on a given surface for the ray to be retroreflected. \par
To stringently prove this, we consider the CCR of Fig. \ref{fig:system_design}. Rays are incident at an angle described by the propagation vector $k_0 = ( \alpha, \beta , \gamma )$, where normal incidence corresponds to $\alpha = \beta = \gamma = -\frac{1}{\sqrt{3}}$. For rays hitting the sensor (on facet A) on the first bounce, the angle of incidence is shown below.
\begin{equation}
   \theta_A = \cos^{-1}\left(\frac{k_0 \cdot -\hat{n}_A}{|k_0|}\right) =
   \cos^{-1} \left( \frac{\gamma}{\sqrt{\alpha^2 + \beta^2 + \gamma^2}} \right) 
\end{equation}
Rays hitting the sensor on the second bounce incident, can come from face B or C, in which case the propagation vectors are $(-\alpha, \beta, -\gamma)$ and $(\alpha,-\beta,-\gamma)$ respectively. Therefore, the angle of incidence to the sensor located on facet A is:
\begin{equation}
   \theta_A = \cos^{-1}\left(\frac{k_1 \cdot -\hat{n}_A}{|k_1|}\right) =
   \cos^{-1} \left( \frac{\gamma}{\sqrt{\alpha^2 + \beta^2 + \gamma^2}} \right) 
\end{equation}
The important observation is that the sensor has the surface normal $\hat{n}_A = (0, 0,-1)$, and this means that the rays hitting the sensor on the first and second bounce have the same incident angle on the sensor. This argument can be extended to show that the third bounce also has the same incident angle on the sensor. If we normalize the propagation vectors, the incident angle on the sensor is simply $ \theta_A = \cos^{-1}(\gamma) $. We will consider incident vectors normalized hereafter.

\subsection{Ray Tracing for the Near Field Pattern}
In order to calculate the near field pattern, we identify the regions of the light exiting the corner cube that have interacted with the sensor. To do this, we break down the problem into sets of rays bounded by polygons that have the same propagation vector. We track these polygons as they bounce between the three surfaces of the corner cube and exit. For retroreflection to occur, a ray must hit all three surfaces before exiting the corner cube, therefore any ray that does not hit all three surfaces is lost and not considered to return to the transceiver.
We begin by defining the finite surfaces of the corner cube to form three squares of unit side-length on the XY, YZ, and XZ planes for facet A, B and C respectively, with the center of the corner cube being (0,0,0). This is shown in Fig. \ref{fig:system_design}. We keep track of the direction of propagation starting by defining the incident propagation vector. We define an incoming ray at normal incidence to the corner cube as having a propagation vector $\alpha = \beta = \gamma = \frac{-1}{\sqrt{3}}$ such that the incident ray is normal to the aperture of the corner cube formed by the plane containing (0 1 1), (1 0 1), and (1 1 0).

\begin{figure}
    \centering
    \includegraphics[width=\linewidth]{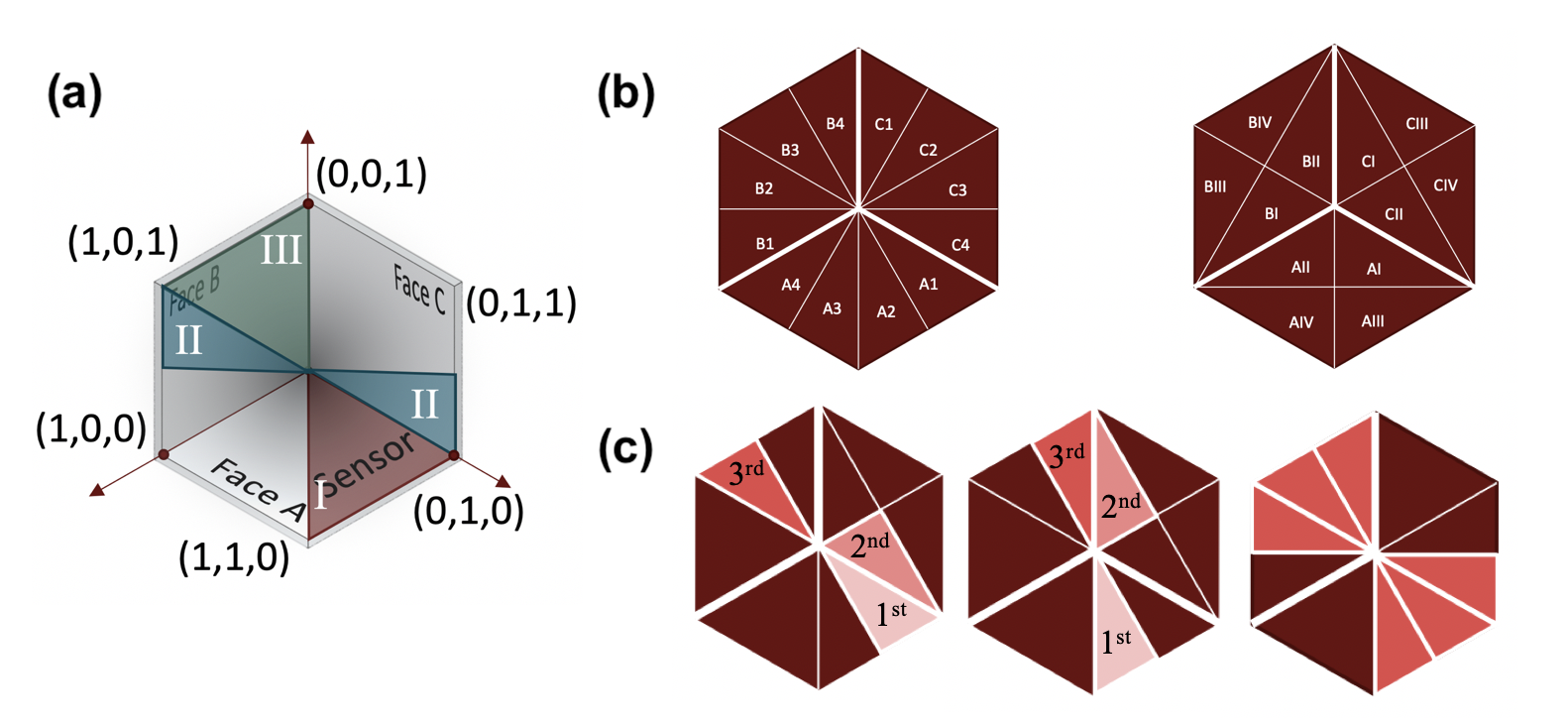}
    \caption{Labeled corner reflectors with named triangular faces. (a) the regions show light exiting the CCR that hit the sensor on the numbered bounce (green-3rd, blue-2nd, red-1st). (b) naming scheme for describing bounces at normal incidence. (c) demonstration of the two unique cases of light incident a triangle in the CCR propagating through the device.}
    \label{fig:intuition}
\end{figure}
We first consider the simple, but important, case of normal incidence, to build an intuition for how the device works. Using the naming scheme of Fig. \ref{fig:intuition}b and \ref{fig:intuition}c, light incident on A1 will bounce onto surface CII before exiting B3. The sensor is located on surface A1 (and A2), so light exiting B3 will have interacted with the sensor. By symmetry, the same argument is made for C4 which has an intermediate bounce on A1 thus interacting with the sensor and coming out at B2. In the same way, light incident on A2 will interact with the sensor, have an intermediate bounce on CI and exit on B4. By reciprocity light exiting A1, B3, C4, B2, A2, and B4 will all have interacted with the sensor. All other triangles in the naming scheme of Fig. \ref{fig:intuition}b,c will not have interacted with the sensor. Thus 6 of 12 triangles, or half of the returning beam, will have the sensor information encoded in it, and half will not, as shown in the rightmost panel of Fig. \ref{fig:intuition}c. \par
This special case illustrates the advantage of placing the sensor on one half of one facet of the CCR: If we have a phase sensor with a \(\pi\) phase shift, the near field and therefore the far field is completely changed. The symmetry has changed from hexagonal to very nearly square, and the equal near-field areas with opposite phase, lead to a null in the center of the far field. The change from hexagonal to (close to) square symmetry and the appearance of a null in the center of the diffraction pattern are illustrated in Figs. \ref{fig:normalincidence_phases} and \ref{fig:3d_normal} below. \par
For the general case of arbitrary angles of incidence, we find the near field by using a 3D ray tracing algorithm that is comprised of three basic functions: reflection, propagation, and intersection. With these three functions, we are able to track every segment of light that exits the corner cube on a retroreflective path. We keep track of the points that bound the polygons of rays in matrices $S_i$ and the corresponding propagation vectors for those matrices as $k_i$.\par
The reflection operation is performed by reflecting the k vector of the facet defined by the surface normal vector ($n_F$). The reflection matrix is generated from equation \ref{eq:reflection} where $I$ is an identity matrix. By multiplying $n_F$ by its transpose, we get a 3x3 matrix with only one non-zero diagonal element of unit value. When multiplied by 2 and subtracted from the identity matrix I as described in equation \ref{eq:reflection}, it forms a reflection matrix.

\begin{equation} 
\mathbf{R}(\hat n_F ) = I - 2\ \hat n_F \cdot \hat n_F\ ^T
\label{eq:reflection}
\end{equation}
The reflection matrices for facets A, B, and C are shown below where the non-zero component in the facet normal vector causes a negation of that element of the propagation vector. This operator is a definitional reflection about the normal vector to the plane. \par

\begin{equation}
\scalebox{0.85}{$
 R(\hat{n}_A) =    
\begin{bmatrix}
1 & 0 & 0\\
0 & 1 & 0\\
0 & 0 &-1
\end{bmatrix}
 R(\hat{n}_B) =    
\begin{bmatrix}
1 & 0 & 0\\
0 & -1 & 0\\
0 & 0 &1
\end{bmatrix}
 R(\hat{n}_C) =    
\begin{bmatrix}
-1 & 0 & 0\\
0 & 1 & 0\\
0 & 0 &1
\end{bmatrix} $}
\end{equation}
The reflected propagation vector is the multiplication of the incoming propagation vector with the rotation matrix.
\begin{equation}
    \hat{k}_{n+1} = R(\hat{n}_F)\hat{k}_n
\end{equation}
Propagating rays to the next surface is approached by a matrix method wherein we project the set of points $S_F$ on one facet ($F$) with a corresponding propagation vector ($k_n$) to the next facet $F'$. We represent sets of rays as bounded triangles of rays that share a propagation vector. In general, a projection can cause rays to fall outside of the finite facets of the corner cube and cause the set of bounding points to change topologically, for instance from a triangle to a quadrilateral. We thus define a propagation function that takes an arbitrary number of bounding points and make $S_F$ an Mx3 matrix where each of the M bounding points is a row vector.\par
We project each bounding point in $S_F$ along the shared propagation vector $k_n$. We use the normal vector of the next facet $n_F'$ to determine the position for each bounding point in the plane of the next facet of intersection. This leads to the following equation for the next set of intersection points (see supplemental A for the derivation of equation \ref{eq:projection}).
\begin{equation}
\label{eq:projection}
    \mathbf{S_{F'}^{\infty}} = Proj_{F \rightarrow F'}(\mathbf{S_{F'}} )= -\left( \frac{\mathbf{S_F}\ \cdot \ \hat{n_{F'}}}{\hat{k}^T\ \cdot \ \hat{n_{F'}}}\right) \hat{k}^T+ \mathbf{S_F}
\end{equation}

After projection, some of the projected points on the next surface may fall outside the next facet of the CCR. These rays will not retroreflect. We define the facet of the corner cube $F_F$ to be the four bounding points of that facet. To determine what part of the projection that fall outside the facet, we define a function that finds the intersection of the set of rays $S_F$ with the facet $F_F$ of the CCR. This function is conducted with python shapely’s function "intersect", which is based on the classic numerical intersection of convex polygons \cite{geometric_intersection}. The intersection of convex polygons results in a convex polygon, so this approach can be repeated for the third reflection. Thus we can describe the intersection generically with equation \ref{eq:intersection}.

\begin{equation}
\label{eq:intersection}
    \mathbf{S_{F'}}= \mathbf{S_{F'}^\infty}\cap \mathbf{F_{F'}}
\end{equation}
These three functions, reflection, projection, and intersection, combine to a practical and efficient numerical calculation of the near field. \par
Our objective is to calculate regions \textbf{\textit{T}} and \textbf{\textit{N}} in Eq. \ref{eq:diff_integral} which correspond to the exit aperture shapes for the regions that have and have not undergone sensor interaction. Retroreflected light hit all three facets in one of six permutations: ABC, ACB, BAC, BCA, CAB, CBA. (As shown in Figs. \ref{fig:system_design} and \ref{fig:intuition}, the sensor is on facet A). Of these six cases, in two of them the light hit the facet containing the sensor (A) first, in two of them the light hits A on the second reflection, and in two the light hit A last. Due to reciprocity of electromagnetic wave propagation in non-magnetic materials, the four permutations where the light hits the sensor first and last can be treated the same way, while the two cases where the light hits A on the second reflection require slightly different treatments. 
For the case where the sensor is hit on the first bounce, we use the operations of reflection, propagation, and intersection to trace the light incident on the sensor triangle to its output polygon. By reciprocity, we know that light incident on the output polygon will exit on the sensor triangle, so the output polygon and the sensor triangle are two parts of the sensor region (\textbf{\textit{T}}) in Eq. \ref{eq:diff_integral}. Likewise, we trace the non-sensor triangle on facet A to is output polygon and we get two parts of the non-sensor output (\textbf{\textit{N}}). The two permutations BAC and CAB are treated the same way taking the intersection with the sensor region on the second bounce to get the full regions \textbf{\textit{T}} and \textbf{\textit{N}}. 
To complete the calculation of the near field, we rotate the reference frame such that the initial propagation vector is aligned with the optical axis -z, and we obtain a two-dimensional sensor aperture along our optical axis that represents our near field sensor pattern. (See supplemental B for details).

\section{Results}
\subsection{Diffraction Patterns}
\begin{figure}[H]
    \centering\includegraphics[width=.9\linewidth]{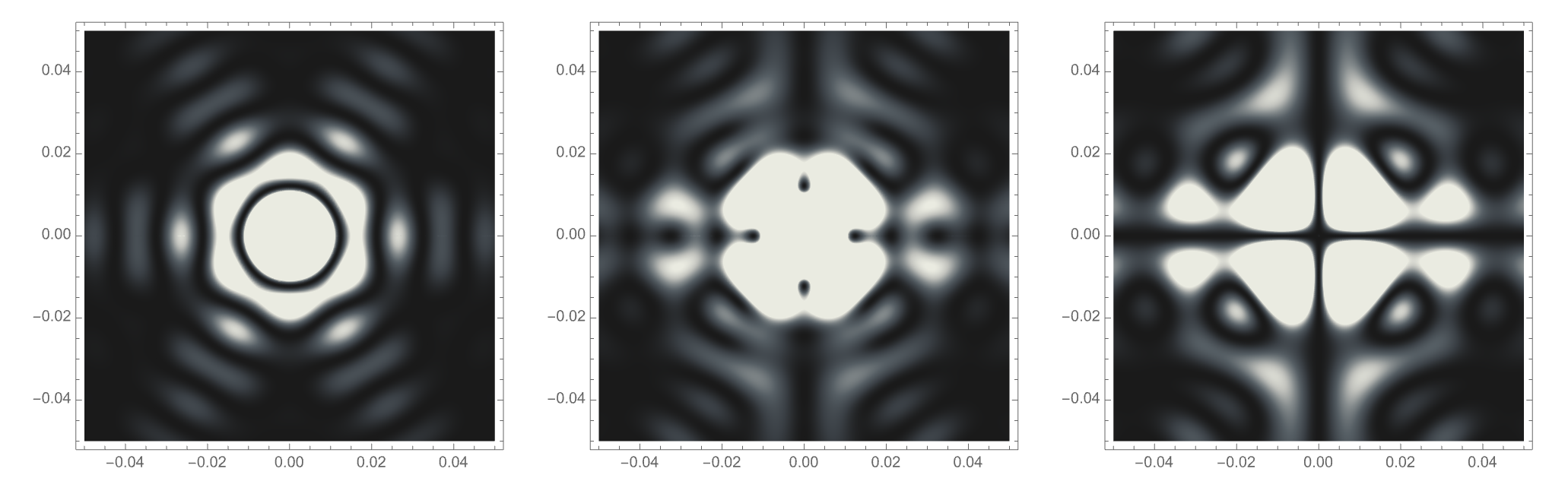}
    \caption{Diffraction patterns with phase variation at normal incidence - from left to right diffraction patterns from device with sensor phase values of 0, $\pi/2$, and $\pi$}
    \label{fig:normalincidence_phases}
\end{figure}

\begin{figure}[H]
    \centering
    \includegraphics[width=0.45\linewidth]{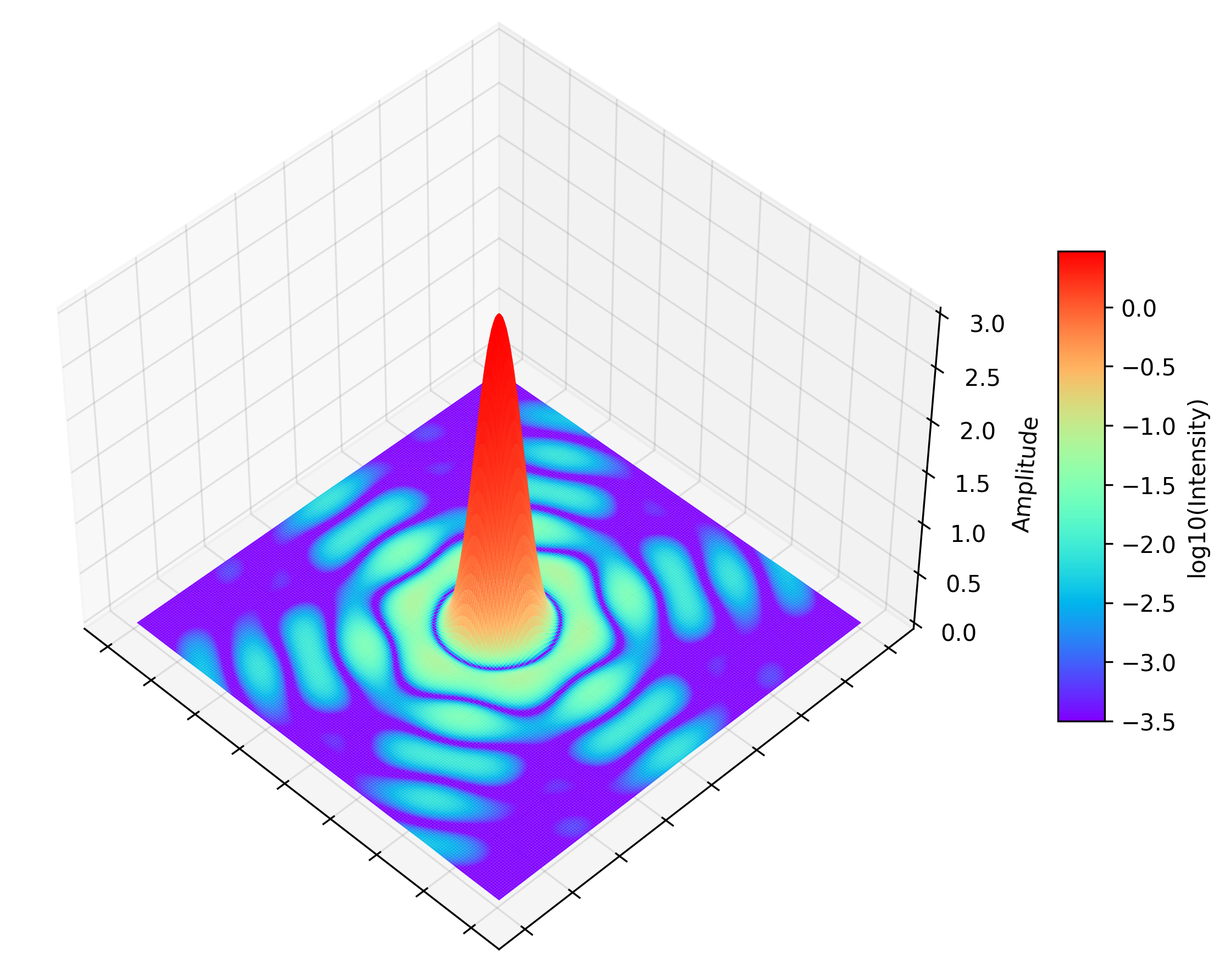}
    \includegraphics[width=0.45\linewidth]{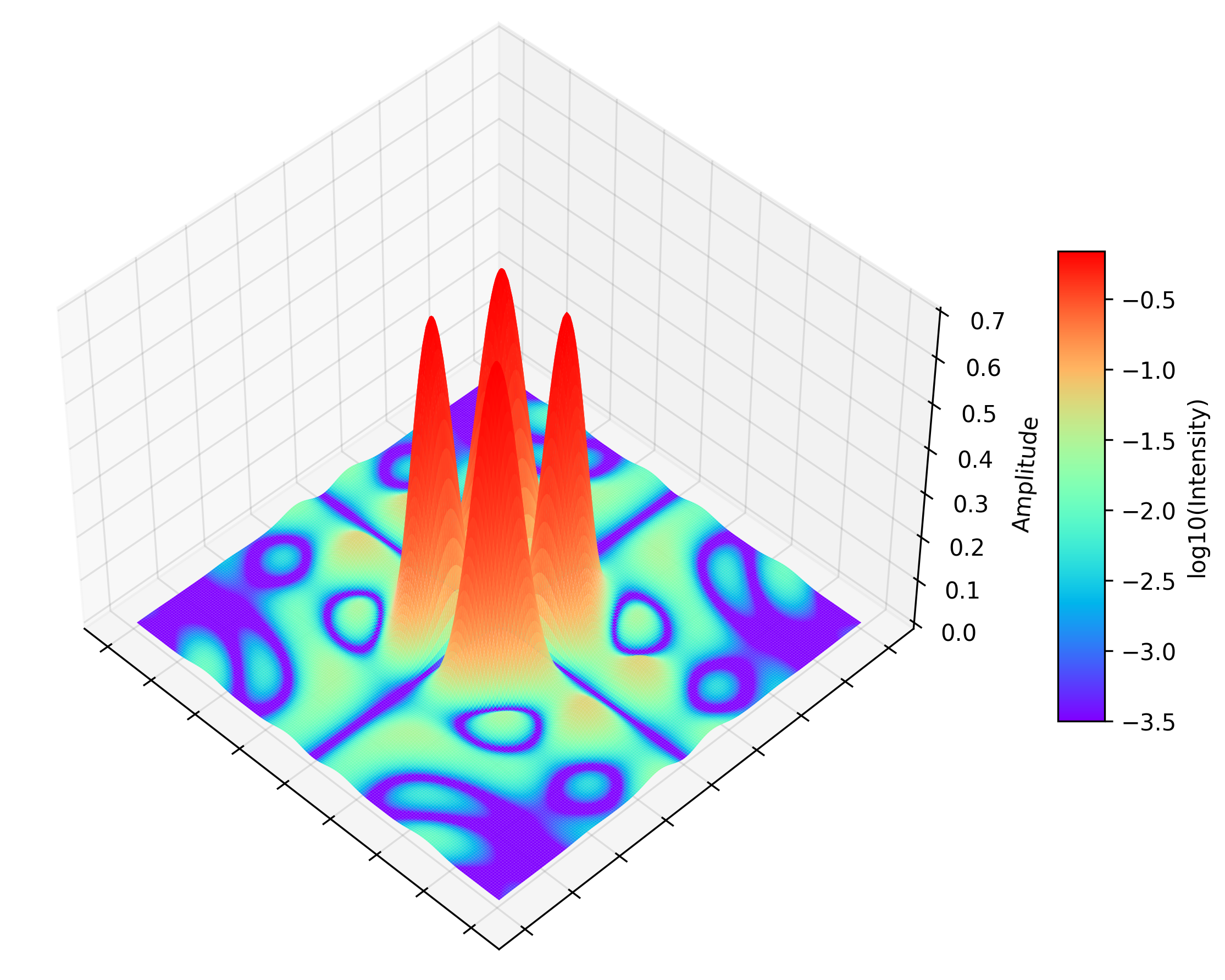}
    \caption{Three dimensional visualization of the diffraction pattern with linear height and logarithmic colors visualizing patterns for 0 phase shift (left) and \(\pi\) phase shift (right)}
    \label{fig:3d_normal}
\end{figure}

We use the described approach to calculate analytical descriptions of diffraction patterns across the parameter space. As the sensor's phase is modulated from 0 to $\pi$ relative to the reference surface, the symmetry of the diffraction pattern moves from having 6-fold radial symmetry to approaching four-fold radial symmetry at normal incidence as seen in Figs. \ref{fig:normalincidence_phases} and \ref{fig:3d_normal}. This symmetry change is intuitively explained by the near field pattern symmetry shown in Fig. \ref{fig:intuition}. This sensor aperture superimposes a 4-fold symmetry on the hexagonal structure resulting in symmetry changes, particularly when the phase is altered such that destructive interference occurs in the 0\textsuperscript{th} diffraction order. 

\begin{figure}[H]
    \centering\includegraphics[width=.9\linewidth]{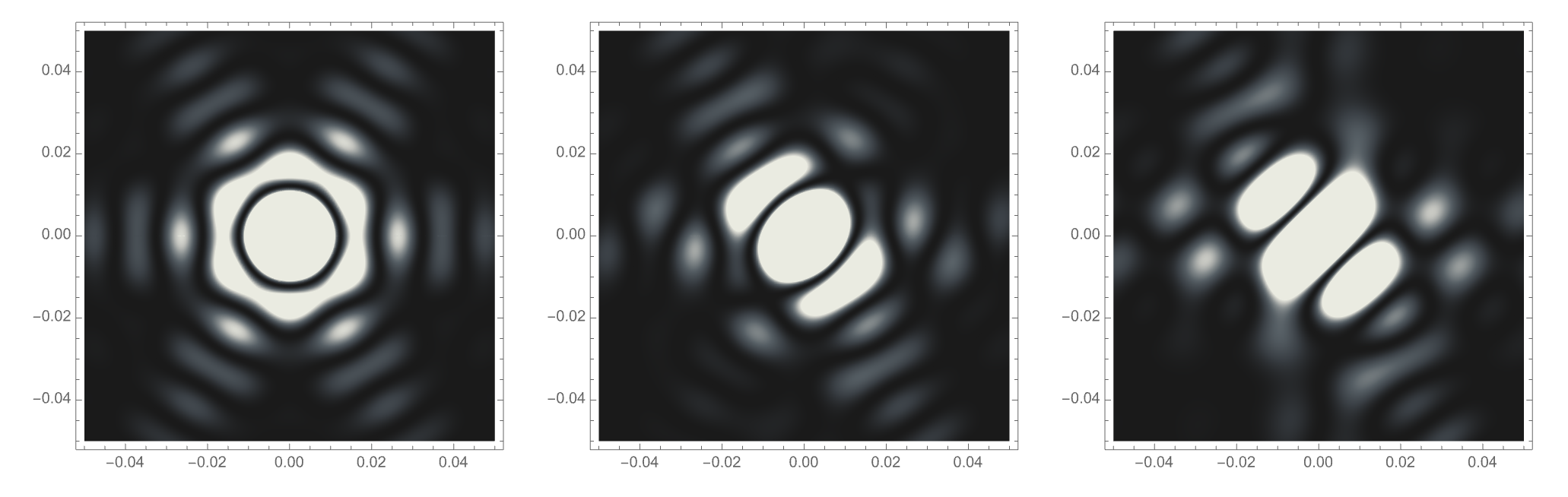}
    \caption{Diffraction patterns with amplitude variation at normal incidence - from left to right diffraction patterns from device with sensor reflection of values of 1, 1/2, and 0 relative to the reference mirror surface}
    \label{fig:normalincidence_amps}
\end{figure}

A change in the sensor amplitude imparts a different reduction in the symmetry as seen in \ref{fig:normalincidence_amps} due to the lack of destructive interference.

When the sensor surface is equal in phase and reflected amplitude to the reference surface, we get the diffraction pattern seen in figures \ref{fig:normalincidence_phases} and \ref{fig:normalincidence_amps} on the left side.
\subsection{Phase Sensitivity}

In this section, we consider phase sensors and resulting changes in diffraction patterns induced by purely phase perturbations. At normal incidence, the 0 phase shift mode and the $\pi$ phase shift mode have orthogonal near fields and far fields (diffraction patterns). We determine this by demonstrating that the power of the two modes overlap integral is zero as shown in equation \ref{eqn:overlapzero}.
\begin{equation}\Big| \oint_{-\infty}^{\infty}D(X,Y,1,0)D(X,Y,1,\pi) dX dY\Big|^2  = 0
\label{eqn:overlapzero}
\end{equation}

\begin{figure}[H]
    \centering
    \includegraphics[width=.9\linewidth]{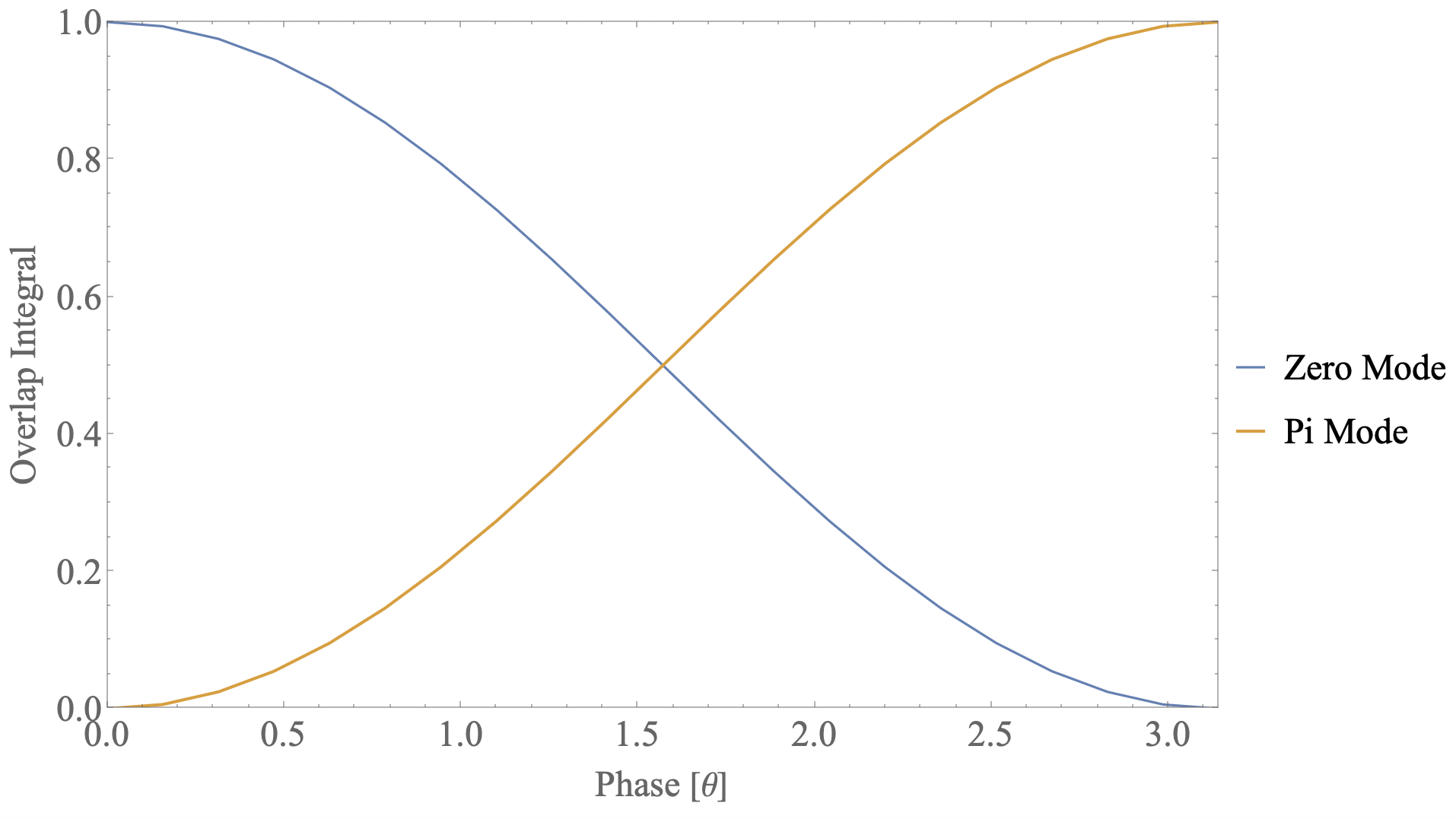}
    \caption{Overlap integral of diffraction pattern with a $\theta$ phase shift with the 0 phase shift mode and the $\pi$ phase shift mode at normal incidence}
    \label{fig:norm_sense}
\end{figure}

At normal incidence the diffraction pattern is therefore a linear combination of the 0 and \(\pi\) diffraction patterns. By calculating the overlap integral of a diffraction pattern with phase shift $\theta$ with the 0 and $\pi$ modes of the phase sensor case in Fig. \ref{fig:normalincidence_phases}, we see that the distribution varies sinusoidally as shown in Fig. \ref{fig:norm_sense}. This behavior matches that of a first order interferometer. 

\begin{figure}[H]
    \centering
    \includegraphics[width=.9\linewidth]{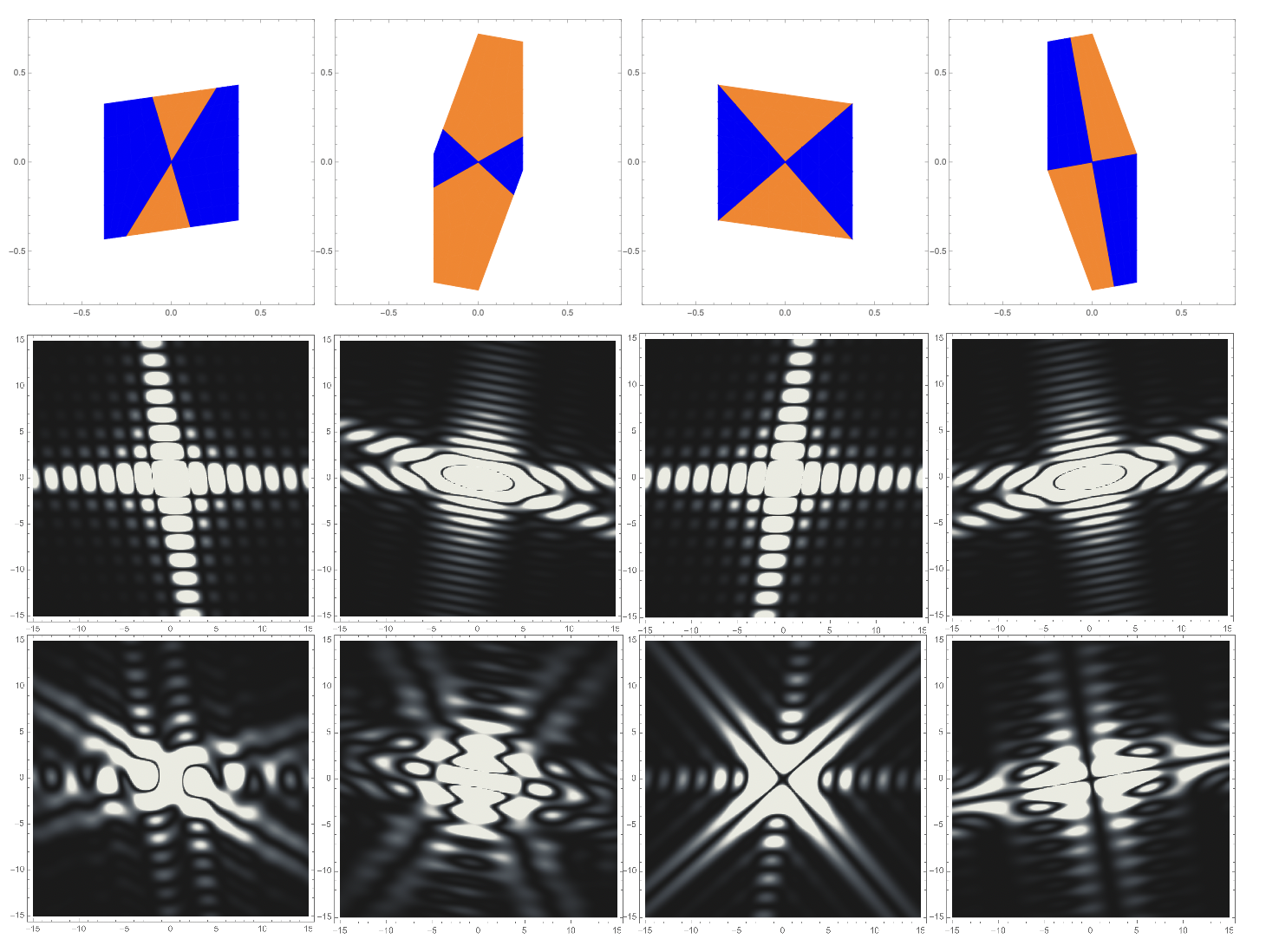}
    \caption{The top row shows near field patterns, the middle row shows the 0 phase shift and the bottom row shows $\pi$ phase shift for 25 degrees off normal rotations.}
    \label{fig:manyAngles}
\end{figure}

At non-normal incidence, the situation is more complex. Varying the angle of incidence alters the diffraction pattern in a way distinct from changes in the sensor’s amplitude or phase. In the near field, this manifests as a reduction of the illuminated aperture with increasing tilt, as shown in Fig.~\ref{fig:manyAngles}.  

To generalize Eq.~\ref{eqn:overlapzero}, we define the normalized overlap integral between the zero-phase diffraction field \(D(X,Y,1,0)\) and a phase-shifted field \(D(X,Y,1,\theta)\) as
\begin{equation}
L(\theta,\xi,\eta) = 
\frac{\displaystyle \iint_{\mathcal{A}} D(X,Y,1,0;\xi,\eta)\, D(X,Y,1,\theta;\xi,\eta)\, dX\,dY}
{\displaystyle \iint_{\mathcal{A}} |D(X,Y,1,0;0,0)|^2\, dX\,dY},
\label{eqn:overlapnorm}
\end{equation}
where the integration is performed over the fixed receiving aperture \(\mathcal{A}\), and \(\xi,\eta\) denote the two tilt angles that determine the angle of incidence and direction of incidence of the incoming light to the corner cube. Normalization to the zero-phase, normal-incidence field makes clearer how tilt modifies both the aperture and the balance between sensor and reference regions.\par

\begin{figure}[H]
    \centering
    \begin{subfigure}[t]{0.2\textwidth}
        \centering
        \caption*{\textbf{(a)}}
        \includegraphics[width=\linewidth]{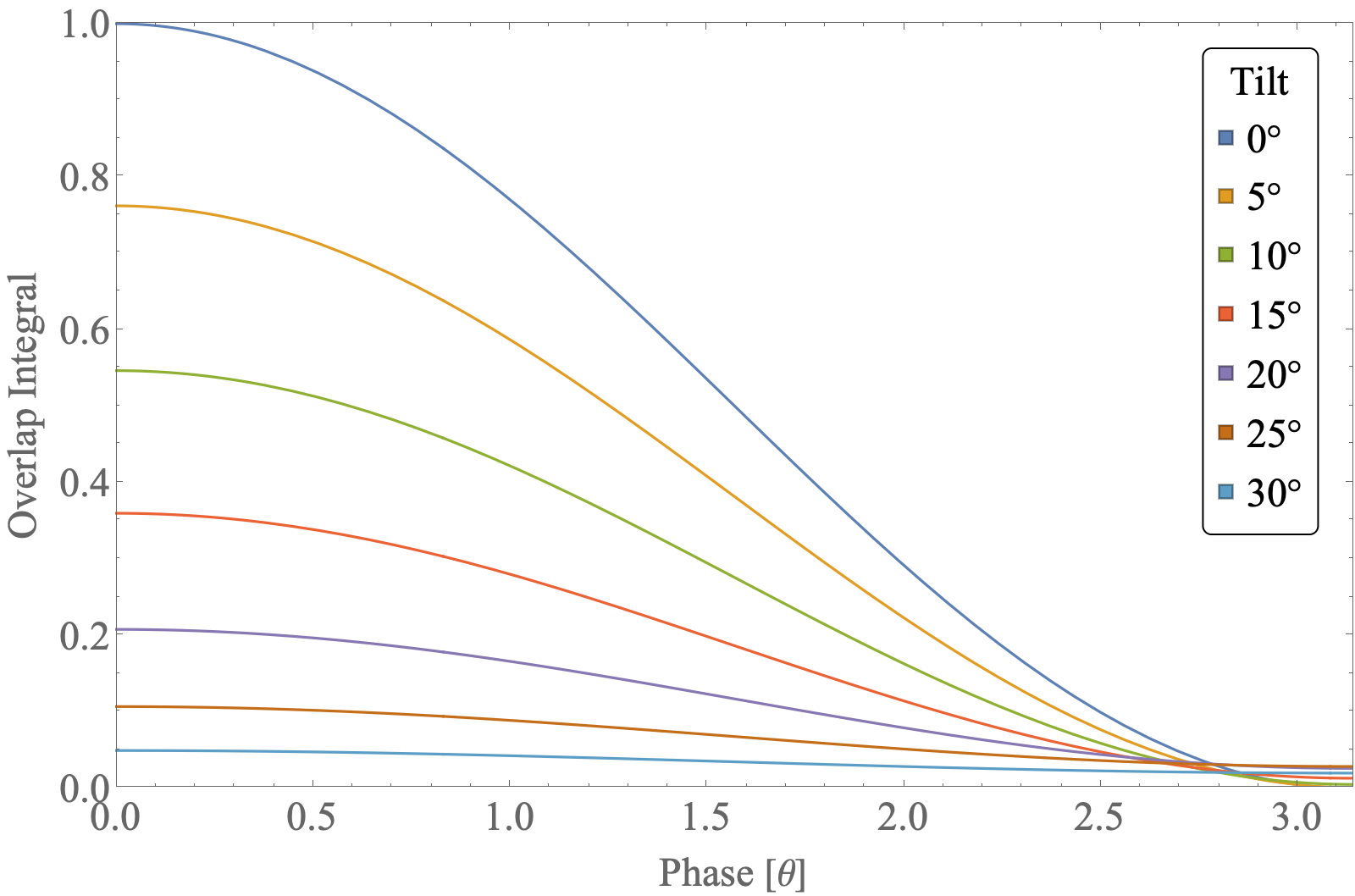}
        
    \end{subfigure}
    \begin{subfigure}[t]{0.2\textwidth}
        \centering
        \caption*{\textbf{(b)}}
        \includegraphics[width=\linewidth]{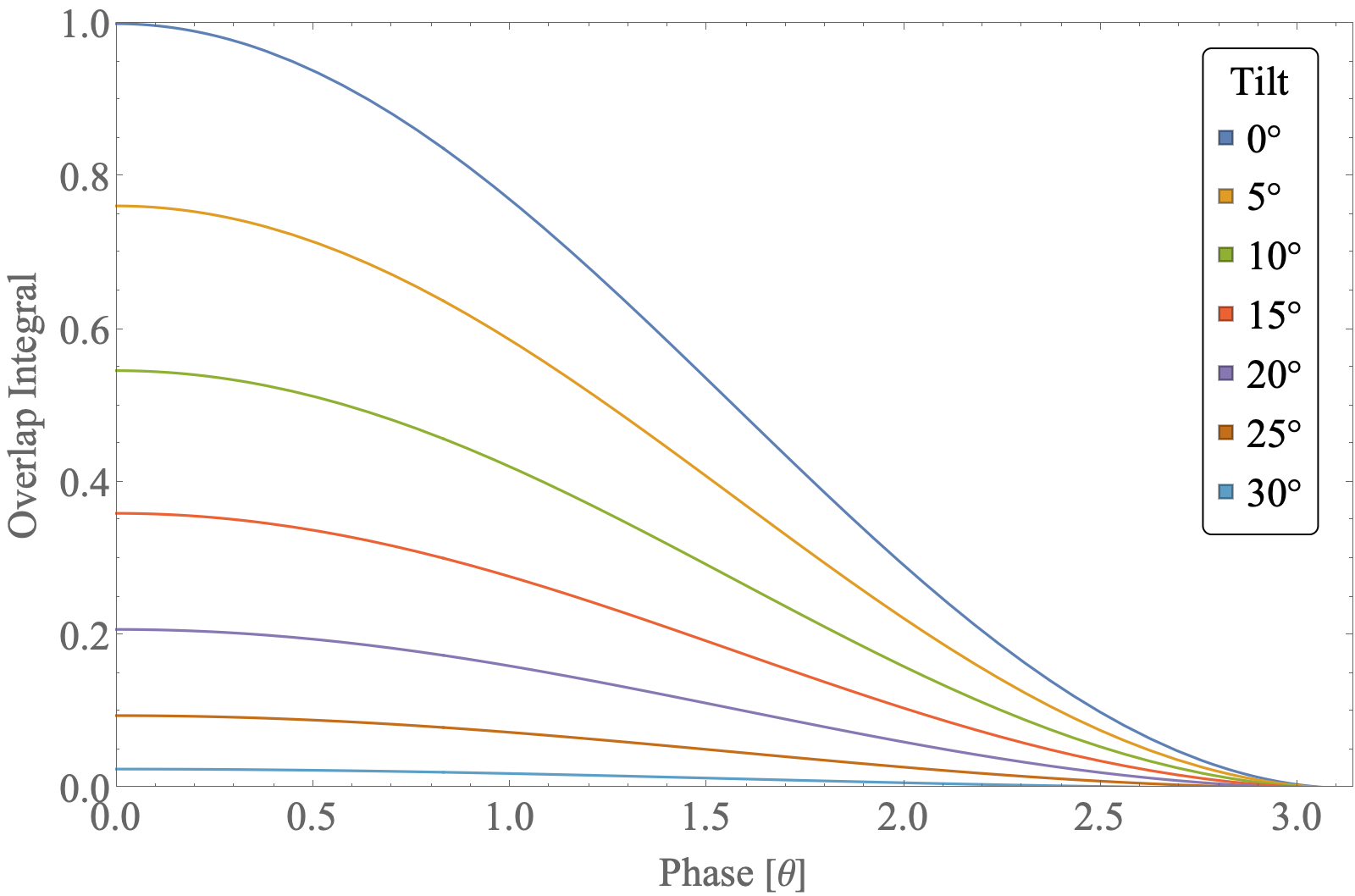}
        
    \end{subfigure}
    \caption{Overlap integrals of 0 phase shift mode with phase shifted mode at different incidence angles for a) Asymmetric case — rotation towards (0,0,-1) and b) Symmetric case — rotation towards (0,1,0).}
    \label{fig:angle_overlap}
\end{figure}

As the corner cube is rotated, the effective aperture shrinks, reducing the overall magnitude and thereby decreasing the slope of the overlap integral and thus the phase sensitivity. Along some directions of rotation, the ratio of light in the near-field that has interacted with the sensor versus that which has interacted with the reference regions remains equal, and the overlap integral goes to zero at a phase shift of \(\pi\) (Fig.~\ref{fig:angle_overlap}b). We consider these rotations to be the symmetric cases, producing near fields and diffraction patterns seen in the right six panels of Figure \ref{fig:manyAngles}. Other directions of rotation result in this ratio deviating from unity, producing an imbalance in the interferometer that prevents the overlap from vanishing at \(\pi\) (Fig.~\ref{fig:angle_overlap}a). These are the asymmetric cases, seen on the left of Fig. \ref{fig:manyAngles}. The reduced aperture lowers the peak value of the overlap near zero phase, while imbalance alters its minimum, together defining how different angular rotations impact sensitivity.
\par

Figure \ref{fig:angle_overlap} delineates the practical range over which the inverse problem can be expected to yield reliable solutions. At small tilts, the reduction in power is modest and the derivative of the overlap with respect to phase remains large, ensuring sensitivity to phase perturbations. As the incidence angle approaches 30\(^\circ\), the total collected power falls near its useful limit. At that angle, the slope of the overlap integral remains sufficient to support inversion with the exception of near the 0 and \(\pi\) cases. For the state of the sensor, we determine the useful, invertible region to be between \(\pi/6\) and \(5\pi/6\). Outside of these bounds, both angle and phase sensitivity degrade substantially, so we use these operational boundaries for reconstructing phase from measured diffraction patterns.

In Fig. \ref{fig:angle_overlap} we see that in the imbalanced case in \ref{fig:angle_overlap}a, the change of slope with angle is driven primarily by the reduction in intensity as opposed to the introduced imbalance. This is advantageous because it demonstrates that the imbalance is a weak effect that is minimized when the power is maximized. This yields a system that is conducive to practical implementation as a power maximization as a function of transceiver angle yields a balancing of the arms of the interferometer and a sensitivity maximum.

\subsection{Inverse Estimator Design and Optimization}

In the preceding sections, we described the forward calculation of diffraction patterns from specified incident angles and sensor phase settings, we now consider the inverse problem for phase sensing.

While we use two angles to calculate the various diffraction patterns, we aim to use this inverse solver to understand phase sensors. Typical optical phase sensors change their response not only with a change in the measurand but also the angle of incidence of the light to the surface. This is because the primary response mechanism is a change in the effective thickness or the refractive index of that phase sensor, which directly modifies the accumulated optical phase as a function of illumination angle. Therefore, while we use the angles of incidence to the corner cube to generate diffraction patterns, we condition the output to measure the incidence angle of light to the sensor surface along with the sensor phase. The combination of an accurate estimation of the incidence angle and the sensor phase, provides a direct route to predicting refractive or effective index of different types of optical sensors (resonators, thin films, Fabry-Perot etc).

We demonstrate an algorithm capable of converging to the global minimum across the optimization space across the parameters of interest (angle of incidence to the sensor and sensor phase). The convergence of the algorithm supports the claim that there are no troublesome degeneracies in the system. To this end, we implement a three-stage optimization approach that numerically demonstrates global convergence across a bounded space. As discussed, based on the information in Fig. \ref{fig:angle_overlap}, we bound the space with angular inputs within 30 degrees of normal incidence(resulting in an angle to sensor to be from 28 to 80 degrees) and phases between \(\pi\)/6 and 5\(\pi\)/6.

Our inverse problem estimator begins with a lookup table (LUT) of 23805 precomputed diffraction patterns equally angularly separated across the space. Specifically, a \(23\times23\) grid of incidence directions was sampled uniformly in tilt about the x and y axes, spanning angles from \(-30^\circ\) to \(30^\circ\). This construction yields approximately uniform angular spacing of about \(2.6^\circ\) within a \(30^\circ\) spherical region centered on normal incidence to the corner cube. For each point on this grid, diffraction patterns are generated for 45 phase values between \(\pi\)/6 and 5\(\pi\)/6, giving a phase grid of about 0.07 radians.\par
This LUT provides a good starting condition for gradient-based refinement. We use this parameter initialization approach because the overall optimization landscape is not convex. Local minima arise from the lobed structure of diffraction images, where alignment and misalignment of different diffraction patterns cause the image similarity metric to vary nonlinearly as a function of the output parameters. Direct gradient descent from arbitrary initialization is therefore prone to slow convergence and entrapment in these inaccurate minima.
\begin{figure}[H]
    \centering
    \includegraphics[width=\linewidth]{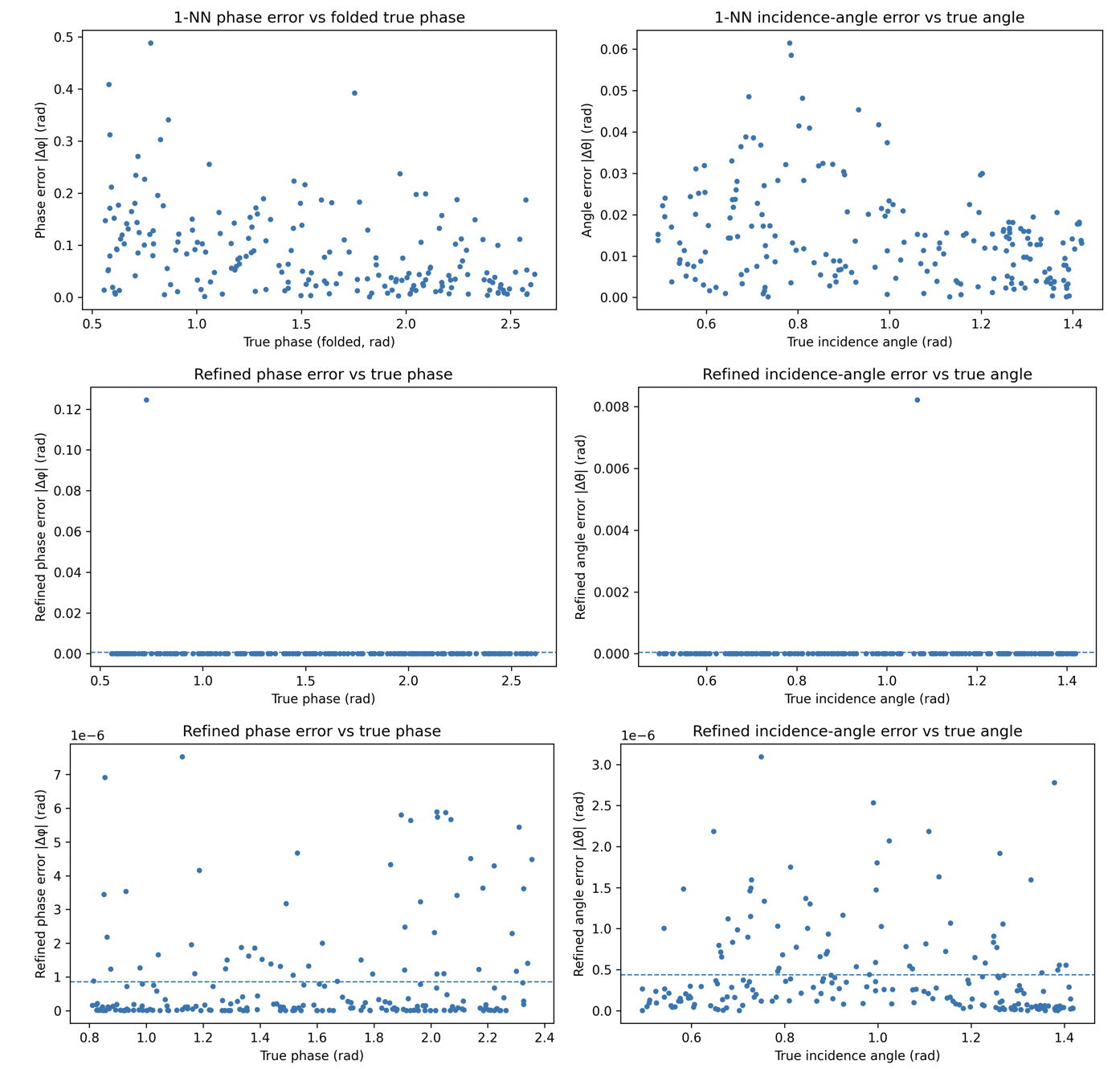}
    \caption{Phase and Angle output error as a function of input parameters after the three steps of our lookup table and LM algorithm. The figures at the top show $10^{-2}$ accuracies of the lookup table for angle and $10^{-1}$ for phase. The middle figures show first pass LM that indicate the many success cases with outlier cases that need to be revisited. The final figure shows the converged data obtaining $10^{-6}$ convergence for both phase and angle.}
    \label{fig:convergence}
\end{figure}

Following LUT initialization, we employ a Levenberg--Marquardt (LM) algorithm \cite{dennis-schnabel-ch10}, a damped least-squares method well suited for nonlinear least-squares problems such as those with sinusoidal structure. LM combines gradient descent with the Gauss-Newton algorithm (GNA) to consider the gradient and an approximation of the curvature when determining the size of the optimization step and defining convergence. Our procedure applies LM in two stages. The first stage uses higher weights, allowing the optimizer to traverse the parameter space more aggressively and escape shallow local minima. Once the residual loss between simulated and measured images falls below a threshold, we reduce the weights to promote smoother convergence and prevent overshooting during the final refinement.

We evaluated this approach by testing the convergence of 200 randomly generated initial points within the parameter space. Of these, 99.5\% (199 trials) converged successfully to a residual image difference loss at or below the order of $10^{-6}$. The remaining 1 trial terminated prematurely at higher residuals (approximately $10^{-3}$), corresponding to time-outs before convergence. For these instances that do not converge, we begin the LM algorithm again from a starting point that is dithered from the LUT nearest neighbor point. Our algorithm adds +/- 5\% to each of the three parameters total range and goes through the LM algorithm for each dithered starting point and checks for convergence. This was effective for getting convergence, and as such we believe this lack of convergence is due to some LUT points falling near local minima that hamper convergence very occasionally. Importantly, the points that do not initially converge have considerable image differences such that the algorithm is certain that they have not converged indicating that these cases are not inherent degeneracies but instead convergence challenges with our LM algorithm.

These results show that phase and angle of incidence can be treated as separable parameters, indicating that phase information can be encoded and retrieved from the diffraction patterns of the modified corner cube.

The success of our algorithm in determining phase and angle from diffraction patterns from the modified corner cubes, demonstrates that the phase reconstruction is convergent across the parameter space and enables deterministic recovery without supervised learning or heuristic pattern recognition. While not used in this demonstration, future experimental implementations may benefit from incorporating physics-informed machine learning to improve robustness under the variability of  real, non-ideal corner cubes.

\subsection{Arrays of Corner Cubes}
\begin{figure}[H]
    \centering
    \includegraphics[width=\linewidth]{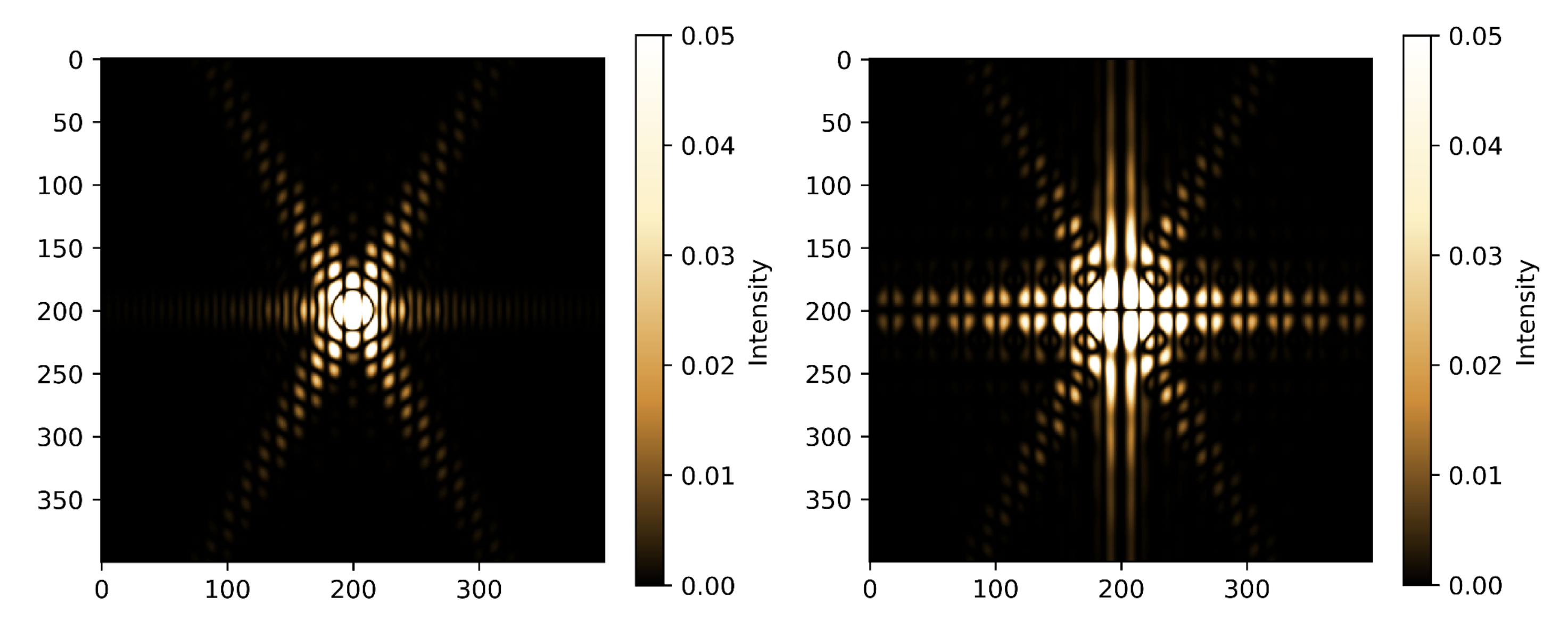}
    \caption{Corner cube array of 2 corner cubes at normal incidence at 0 and \(\pi\) phase shift with sensors on opposing sides of their respective corner cubes.}
    \label{fig:multicube}
\end{figure}

We have discussed the use of a single corner cube for the readout of a single sensor, however these approaches and methods can be used for arrays of corner cubes and collections of sensors. Such arrangements of corner cube arrays create a large variety of diffraction patterns that can be calculated using the techniques described in this paper. A simple example of the diffraction patterns from two adjacent corner cubes are shown in Fig. \ref{fig:multicube}. Extensions to multiple corner cubes and arrays of corner cubes offer additional design degrees of freedom that can be used, for example, to counteract the sensitivity change as a function of angle. This also offers the opportunity to perform multiple sensor readout from diffraction patterns and additional ways of encoding information. 

\section{Discussion}
The modified corner cubes of this paper changes diffraction patterns based on amplitude or phase modulation across half of one facet of the corner cube. Many other arrangements of the modulated area could be beneficial to achieve certain characteristics. An advantage of our approach is that it encodes modulation as a structural feature of the diffraction field. These structural features give us the state of the sensor, but the information encoded in the diffraction pattern exceeds standard interferometers as we can analyze the position of the diffraction orders to extract additional information about the angle of incidence. The sensor does not require matched optical paths or a stable reference arm; the geometry of the corner cube and system architecture provide both a reference and a mixing mechanism, enabling detection with a traditional camera.

Importantly, this method shifts the cost and complexity of sensing from the sensor node to the receiver. Fabrication of the patterned retroreflectors is compatible with inexpensive fabrication processes, and the devices themselves require no onboard power or electronics. All signal interpretation is offloaded to the receiver, decoupling sensing hardware from computational and communication infrastructure.

This architecture addresses key limitations in environmental sensing, where challenges often stem from power, deployment, and communication infrastructure rather than sensor performance. For example, chemically selective thin-film sensors remain difficult to deploy at scale due to packaging and readout constraints. A passive optical platform capable of extracting phase from a return image eliminates many of these barriers. A single scanning transceiver can interrogate multiple corner cubes distributed across the environment, enabling scalable, remote readout without embedded electronics. When fabricated from silicon, these devices can also be made biocompatible and designed to degrade over time, eventually becoming indistinguishable from sand.

We have focused on remote sensing in this paper, but many other applications are readily implemented, including communication and creation of corner cubes with specific signature diffraction patterns.  Patterned corner cubes can function as optically addressable encoders, useful in applications such as localization and encrypted identifiers. Because the devices are retroreflective, they produce bright, directionally confined signals that can be read at long range with minimal background interference. Their optical brightness falls off linearly with distance, in contrast to the quadratic decay of diffusely scattering tags.

\section{Back matter}
\begin{backmatter}
\bmsection{Funding} The authors acknowledge funding from the Stanford Woods Institute for the Environment, the John and Kate Wakerly Stanford Graduate Fellowship, and the National Science Foundation's Graduate Research Fellowships Program.

\bmsection{Acknowledgment} Thanks to Gabriele Cavicchioli, Simon Lorenzo, Carson Valdez, and Anna Miller for their many conversations and support.

\bmsection{Disclosures} The authors declare that a provisional patent application has been filed related to the technology described in this work.

\end{backmatter}
\nocite{*}
\bibliography{ccr_theory_ref}

\begin{thebibliography}{10}
\newcommand{\enquote}[1]{``#1''}

\bibitem{pister_micro_ccr_coms}
D.~Gunawan, L.~Lin, and K.~Pister, \enquote{Micromachined corner cube reflectors as a communication link,} {\protect\JournalTitle{Sensors and Actuators}}  (1995).

\bibitem{olav_modulator_ccr}
D.~Pedersen and O.~Solgaard, \enquote{Free-space communication link using a grating light modulator,} {\protect\JournalTitle{Sensors and Actuators}}  (2000).

\bibitem{scholl_ray_1995}
M.~S. Scholl, \enquote{Ray trace through a corner-cube retroreflector with complex reflection coefficients,} {\protect\JournalTitle{JOSA A}} \textbf{12}, 1589--1592 (1995).

\bibitem{eckhardt_simple_1971}
H.~D. Eckhardt, \enquote{Simple model of corner reflector phenomena,} {\protect\JournalTitle{Applied Optics}} \textbf{10}, 1559--1566 (1971).

\bibitem{geometric_intersection}
M.~I. Shamos and D.~Hoey, \enquote{Geometric intersection problems,} {\protect\JournalTitle{17th Annual Symposium on Foundations of Computer Science}}  (1976).

\bibitem{dennis-schnabel-ch10}
J.~E. Dennis and R.~B. Schnabel, \emph{Numerical Methods for Unconstrained Optimization and Nonlinear Equations} (Prentice-Hall, Englewood Cliffs, NJ, 1983), chap.~10, pp. 218--238.

\bibitem{shao_retro_2018}
S.~Shao, A.~Khreishah, and I.~Khalil, \enquote{{RETRO}: {Retroreflector} {Based} {Visible} {Light} {Indoor} {Localization} for {Real}-time {Tracking} of {IoT} {Devices},} in \emph{{IEEE} {INFOCOM} 2018 - {IEEE} {Conference} on {Computer} {Communications},}  (2018), pp. 1025--1033.

\bibitem{jorgenson_surface_2001}
R.~C. Jorgenson, \enquote{A surface plasmon resonance side active retro-reflecting sensor,} {\protect\JournalTitle{Sensors and Actuators B: Chemical}} \textbf{73}, 236--248 (2001).

\bibitem{ahmed_color-selective_2017}
R.~Ahmed, A.~K. Yetisen, S.~H. Yun, and H.~Butt, \enquote{Color-selective holographic retroreflector array for sensing applications,} {\protect\JournalTitle{Light: Science \& Applications}} \textbf{6}, e16214--e16214 (2017).

\bibitem{waqaskhalid_flexible_2018}
M.~Waqas Khalid, R.~Ahmed, A.~K. Yetisen, and H.~Butt, \enquote{Flexible corner cube retroreflector array for temperature and strain sensing,} {\protect\JournalTitle{RSC Advances}} \textbf{8}, 7588--7598 (2018).

\bibitem{han_retroreflection-based_2022}
Y.~D. Han, K.~R. Kim, K.~W. Lee, and H.~C. Yoon, \enquote{Retroreflection-based optical biosensing: {From} concept to applications,} {\protect\JournalTitle{Biosensors and Bioelectronics}} \textbf{207}, 114202 (2022).

\bibitem{zhou_assembled_2002}
L.~Zhou, K.~Pister, and J.~Kahn, \enquote{Assembled corner-cube retroreflector quadruplet,} in \emph{Technical {Digest}. {MEMS} 2002 {IEEE} {International} {Conference}. {Fifteenth} {IEEE} {International} {Conference} on {Micro} {Electro} {Mechanical} {Systems} ({Cat}. {No}.{02CH37266}),}  (2002), pp. 556--559. ISSN: 1084-6999.

\bibitem{zhou_grating-corner-cube-based_2020}
S.~Zhou, V.~Le, Q.~Mi, and G.~Wu, \enquote{Grating-{Corner}-{Cube}-{Based} {Roll} {Angle} {Sensor},} {\protect\JournalTitle{Sensors}} \textbf{20}, 5524 (2020).

\bibitem{she_pd_2019}
X.~She, Y.~Shen, J.~Wang, and C.~Jin, \enquote{Pd films on soft substrates: a visual, high-contrast and low-cost optical hydrogen sensor,} {\protect\JournalTitle{Light: Science \& Applications}} \textbf{8}, 4 (2019).

\bibitem{han_submillimeter-scale_2022}
M.~Han, X.~Guo, X.~Chen, \emph{et~al.}, \enquote{Submillimeter-scale multimaterial terrestrial robots,} {\protect\JournalTitle{Science Robotics}} \textbf{7}, eabn0602 (2022).

\bibitem{ai_accurate_1992}
C.~Ai and K.~L. Smith, \enquote{Accurate measurement of the dihedral angle of a corner cube,} {\protect\JournalTitle{Applied Optics}} \textbf{31}, 519--527 (1992).

\bibitem{lyu_measuring_2021}
H.~Lyu, L.~Kong, S.~Wang, and M.~Xu, \enquote{Measuring corner cube reflectors through ray tracing of a reflected wavefront,} {\protect\JournalTitle{Applied Optics}} \textbf{60}, 6560--6565 (2021).

\bibitem{murphy_polarization_2013}
T.~W. Murphy and S.~D. Goodrow, \enquote{Polarization and far-field diffraction patterns of total internal reflection corner cubes,} {\protect\JournalTitle{Applied Optics}} \textbf{52}, 117--126 (2013).

\bibitem{kumar_advances_2022}
V.~Kumar, S.~K. Raghuwanshi, and S.~Kumar, \enquote{Advances in {Nanocomposite} {Thin}-{Film}-{Based} {Optical} {Fiber} {Sensors} for {Environmental} {Health} {Monitoring}—{A} {Review},} {\protect\JournalTitle{IEEE Sensors Journal}} \textbf{22}, 14696--14707 (2022).

\bibitem{gillanders_composite_2005}
R.~N. Gillanders, M.~C. Tedford, P.~J. Crilly, and R.~T. Bailey, \enquote{A composite thin film optical sensor for dissolved oxygen in contaminated aqueous environments,} {\protect\JournalTitle{Analytica Chimica Acta}} \textbf{545}, 189--194 (2005).

\bibitem{miao_high-sensitivity_2021}
X.~Miao, L.~Yan, Y.~Wu, and P.~Q. Liu, \enquote{High-sensitivity nanophotonic sensors with passive trapping of analyte molecules in hot spots,} {\protect\JournalTitle{Light: Science \& Applications}} \textbf{10}, 5 (2021).

\bibitem{hong_state---art_2020}
T.~Hong, J.~T. Culp, K.-J. Kim, \emph{et~al.}, \enquote{State-of-the-art of methane sensing materials: {A} review and perspectives,} {\protect\JournalTitle{TrAC Trends in Analytical Chemistry}} \textbf{125}, 115820 (2020).

\bibitem{comert_titanium_2016}
B.~Comert, N.~Akin, M.~Donmez, \emph{et~al.}, \enquote{Titanium {Dioxide} {Thin} {Films} as {Methane} {Gas} {Sensors},} {\protect\JournalTitle{IEEE Sensors Journal}} \textbf{16}, 8890--8896 (2016).

\bibitem{luong_sub-second_2021}
H.~M. Luong, M.~T. Pham, T.~Guin, \emph{et~al.}, \enquote{Sub-second and ppm-level optical sensing of hydrogen using templated control of nano-hydride geometry and composition,} {\protect\JournalTitle{Nature Communications}} \textbf{12}, 2414 (2021).

\bibitem{kay_estimation_1993}
S.~M. Kay, \emph{Fundamentals of Statistical Signal Processing, Volume I: Estimation Theory} (Prentice Hall, 1993).

\end{thebibliography}

\end{document}